\documentclass{article}
\usepackage {graphicx}
\begin{document}
\baselineskip .75cm 
\begin{titlepage}
\title{\bf Is QGP really a liquid ?}       
\author{Vishnu M. Bannur  \\
{\it Department of Physics}, \\  
{\it University of Calicut, Kerala-673 635, India.} }   
\maketitle
\begin{abstract}

Here we address the question regarding the nature of quark gluon plasma (QGP), whether it is a liquid or strongly coupled plasma (SCP), using two different phenomenological models, namely quasi-particle model (qQGP) and strongly coupled quark gluon plasma (SCQGP). First we compare these two models, both of which explains the results of lattice simulation of quantum chromodynamics, as a function of plasma parameter and conclude that the QGP is largely ($T > 1.5 T_c$) SCQGP and only for $T < 1.5 T_c$ it may be a liquid.  
\end{abstract}
\vspace{1cm}
                                                                                
\end{titlepage}
\section{Introduction :}
The non-ideal behavior of QGP near $T= T_c$, observed in lattice simulation (LGT) \cite{bo.1} of quantum chromodynamics (QCD) and relativistic heavy ion collisions (RHICs) experiments, is still a highly debated issue in the study of QGP \cite{qgp.1}. There are various QCD based approaches, perturbative \cite{ka.1}, nonperturbative \cite{br.1} and phenomenological models to explain this problem. QCD based approaches, eventhough questionable due to strong coupling constant near $T = T_c$, was partially successful \cite{ka.1,br.1}. At the same time, phenomenological models were successful to explain, but involves many fitting parameters. For example, largely studied qQGP models \cite{pe.1,qqgp.1} has more than 3 parameters and recently introduced new qQGP \cite{nqgp.1,nqgp.2} has still one parameter. Similarly, strongly coupled QGP \cite{ba.1,ba.2} which is inspired by similar studies in electrodynamic plasma, which is called SCQGP, has 2 fitting  parameters. Both of these phenomenological models, qQGP and SCQGP, fit very well the LGT results, eventhough they are developed based on different philosophy. qQGP is based on the fact that the thermal properties of QGP may be understood by studying a system of non-interacting quasi-partons with effective mass related to collective properties of QGP, instead of a system of real partons with QCD interactions. Whereas in SCQGP, it is assumed that, just like in strongly coupled electrodynamic plasma (SCP), the basic interactions, QCD, is approximated by Coulomb interactions between partons and hence the equation of state (EoS) of SCP is modified for QGP. EoS of SCP was derived by partially analytic and partially numerical methods and parameterized as a function of the plasma parameter, $\Gamma$, which depends on the collective properties of the system. That is, $\Gamma$ is defined as the ratio of average potential energy to the average kinetic energy. 

It should be noted that SCQGP \cite{ba.1,ba.2} is different from popularly known sQGP. sQGP means strongly interacting QGP in the sense that the coupling constant $\alpha_s$ is not very small or weak and hence leads to non-perturbative effects like the existence of hadrons, mostly colored hadrons. It is claimed \cite{sh.1} that pressure due to these hadrons reproduces the pressure seen in LGT simulations. Of course, there is some confusion in using sQGP and SCQGP as discussed in \cite{na.1}. Later, Gelman, Shuryak and Zahed introduced another model cQGP (classical QGP) \cite{sh.2}, based on molecular dynamics simulations, which is similar to SCP or SCQGP, but again claimed to be sQGP. Further, it is speculated that sQGP (or cQGP) might exhibit different phases like gas, liquid and solid as in the theory of SCP. Finally, above sQGP is now widely quoted as liquid QGP and based on the smallness of $\frac{\eta}{s}$ ratio, it is also called perfect liquid, etc. Note that the EoS of cQGP, expanded in terms of $\Gamma$, to the lowest order, is very similar to one used in Ref. \cite{ba.1,ba.2} or SCP. Along with this work, Lin and Shuryak \cite{sh.3} also modelled this system by borrowing similar theory from super symmetry, string theory etc., regarding the evaluation of  $\frac{\eta}{s}$ so on and hot discussions on the bounds of $\frac{\eta}{s}$. But, it should be noted that gas, liquid and solid phases in a system with color charged partons is, in fact, SCQGP, analogous to SCP in plasma physics and not sQGP.  Recently, Chernodub and Zakharov \cite{ch.1}, and Liao and Shuryak \cite{sh.4}, independently, proposed a new picture to the existence of non-ideal nature of QGP near $T = T_c$ in terms of magnetic monopoles etc. and yet another idea to explain non-ideal behavior of QGP and it goes on.  

Here we address an interesting observation that the qQGP and SCQGP, seemingly two different models, both explains the LGT results very well. Of course, both the models are based on the collective  properties of QGP, but the formulations are different. We compare the results of these two models and comment on the liquid nature of QGP. 

In section 2 we first briefly review SCQGP model, results and comment on the liquid nature of QGP. In section 3 a brief review of qQGP and results are presented. In section 4 we compare above two models as a function of the plasma parameter $\Gamma$. Results and conclusions on the comparison of two models and the predictions of the models on the nature of QGP is prsented in section 5.     

\section{SCQGP:} 

SCQGP was first proposed in 1999 \cite{ba.1} and it explained remarkably well the LGT results on gluon plasma and recently, it is also applied to flavored QGP and found to explain the LGT results very well \cite{ba.2}. Here one modifies the EoS of SCP to SCQGP \cite{ba.2} as, 
\begin{equation} e(\Gamma) \equiv \frac{\varepsilon}{\varepsilon_{SB}} 
= 1 + \frac{1}{2.7} u_{ex} (\Gamma) \; , \label{eq:scqgp} \end{equation} 
where $\varepsilon_{SB}$ is the Stefan-Boltzmann gas limit of QGP and $u_{ex} (\Gamma)$, is given by, 
\begin{equation} u_{ex} (\Gamma) = \frac{u_{ex}^{Abe} (\Gamma) + 3 \times 10^3 \, \Gamma^{5.7} 
 u_{ex}^{OCP} (\Gamma) }{1 + 3 \times 10^3 \, \Gamma^{5.7} } \; ,  
  \label{eq:uex} \end{equation}
with    
\begin{equation} u_{ex}^{Abe} (\Gamma) = - \frac{\sqrt{3}}{2} \, \Gamma^{3/2} - 3 \, \Gamma^3 
 \left[ \frac{3}{8} \, \ln (3 \Gamma) + \frac{\gamma}{2} - \frac{1}{3} \right] \; \end{equation} 
 and 
\begin{equation} u_{ex}^{OCP} = - 0.898004 \, \Gamma + 0.96786 \, \Gamma^{1/4} 
      + 0.220703 \, \Gamma^{- 1/4} - 0.86097 \; .  \label{eq:uo} 
      \end{equation}
The excess energy $u_{ex}$ due to non-ideal interactions in above equation is obtained from SCP. SCP, a system of quasi-neutral charged system with Coulomb interaction, is extensively studied and the EoS is parameterized as a function of $\Gamma$ and tested for a wide range of $\Gamma$, $\Gamma < 180$, and different systems like white dwarfs, dusty plasma etc. In QGP also we expect Coulomb like interactions, due to one-gluon-exchange, as in the case of hadron spectroscopy. Contrary to hadron spectroscopy, here in QGP, the confinement interaction between partons are neglected. Hence, SCQGP may also behave like SCP and $\Gamma$ for QCD \cite{ba.1,ba.2} may be written as, 
\begin{equation} \Gamma \equiv \frac{<PE>}{<KE>} = \frac{ g_c \frac{\alpha_s}{r_{av}} }{2 T} 
= \left( \frac{4.4 \,\pi a_f}{3} \right) ^{1/3} \frac{g_c}{2} \alpha_s (T) \; ,  \label{eq:ga}  \end{equation}
where $r_{av}$ may be taken as $(\frac{3}{4 \pi n})^{1/3}$ with density $n \approx 1.1 \,a_f T^3$ and $a_f \equiv (16 + 21 \, n_f /2) \pi ^2 /90$. We choose a phenomenological running coupling constant as, 
\begin{equation} \alpha_s (T) = \frac{6 \pi}{(33-2 n_f) \ln (T/\Lambda_T)}  
\left( 1 - \frac{3 (153 - 19 n_f)}{(33 - 2 n_f)^2} 
\frac{\ln (2 \ln (T/\Lambda_T))}{\ln (T/\Lambda_T)} 
\right)  \;. \label{eq:as} \end{equation}
Of course, many authors \cite{pe.2,th.1,th.2} included a factor $2$ in the definition of $\Gamma$ for QCD plasma to include the magnetic interactions, which amounts to a different value for the fitted parameter $g_c$ in our model.  
\begin{figure}[h]
\centering
\includegraphics[height=8cm,width=12cm]{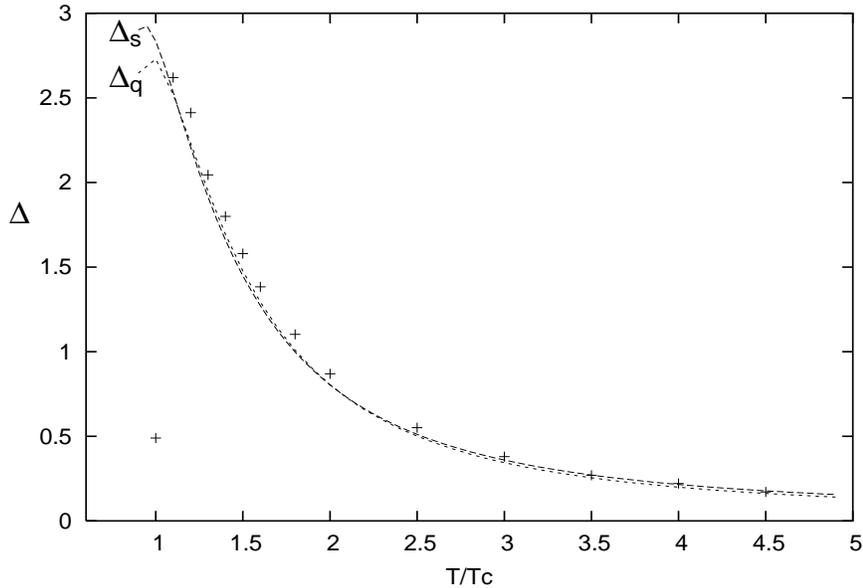}
\caption{ Plot of $\Delta \equiv (\varepsilon - 3 P)/T^4$ as a function of $T/T_c$ for gluon plasma using the models SCQGP ($\Delta_s$), qQGP ($\Delta_q$) and LGT results (symbols) \cite{bo.1}.} 
\end{figure}
In our model $g_c$ and $\Lambda_T$ are the two parameters which may be different for different systems \cite{ba.2} like gluon plasma, 2-flavor, 3-flavor and (2+1)-flavor QGP. Here we consider a simplest system, gluon plasma, and hence $n_f = 0$ in $a_f$ and $\alpha_s (T)$. To fit the LGT results \cite{bo.1}, we used $g_c = 2.8$ and $t_0 \equiv \Lambda_T /T_c = 0.5$ and obtained a very good fit. As an example we plotted $\Delta \equiv (\varepsilon - 3 P)/T^4$ as shown in Fig. 1 ($\Delta_s$) along with that of qQGP model and LGT results \cite{bo.1} and the resulting $\Gamma(T)$ is plotted in Fig. 2. The plasma parameter $\Gamma$ is about $0.5$ at $T = 5 T_c$ and only close to $T = T_c$, it rapidly grows to $1.8$. If we take the results of SCP \cite{ic.1} that plasma is in liquid state for $\Gamma > 1$, QGP is {\it not} in liquid state except very close to $T = T_c$, i.e., for $T_c < T < 1.5 T_c$, in contrary to earlier claim in Ref. \cite{pe.2,th.1,th.2}. Note that this conclusion follows from strongly coupled plasma formalism of QGP which fits LGT results. Even if we take a factor of 2 due to magnetic interactions our conclusions will not change, but the value of fitting parameter $g_c$ will be reduced by a factor of half. It is interesting to note that the plot given in Ref. \cite{pe.2} for $\Gamma(T)$, using qQGP, is similar to Fig. 2 if one scales it by a factor of half and $\Gamma$ is less than $1$ up to $T = 1.5 T_c$ and increases rapidly to greater than $1$ as $T$ decreases. Probably, the liquid-gas phase transition of QGP and confinement-deconfinement phase transition or cross over may be coinsiding and one needs a quantitative description of liquid-gas transition and the critical value of $\Gamma$ ($\Gamma_c$) to fully understand the phenomena. Until then it is too drastic to speculate the existence of liquid and solid or crystal phase of QGP as in \cite{sh.2,th.1} and, from Fig. 2, the maximum value of $\Gamma \approx 1.8$ at $T=T_c$. It is interesting to note that recently Chernodub and Zakharov \cite{ch.1} also have shown that QGP may be in liquid state only for $T_c < T <2 T_c$.        
 
Thus we see that nonrelativistic, classical strongly coupled plasma model with phenomenological modifications to take into account of relativistic, flavor, color factors and quantum effects in terms of running coupling, fits well LGT data in the relevent temperature range, $T_c$ to 5 $T_c$, and shows that QGP is strongly coupled plasma and may not be liquid. 
Of course, this model has a limitations that it is a classical model and hence for weak coupling limit it goes to Debye-Huckel results and not the perturbative QCD results. As shown in Ref. \cite{kp.1}, perturbative results also goes to Debye-Huckel term for classical limit. 
    
\begin{figure}[h]
\centering
\includegraphics[height=8cm,width=12cm]{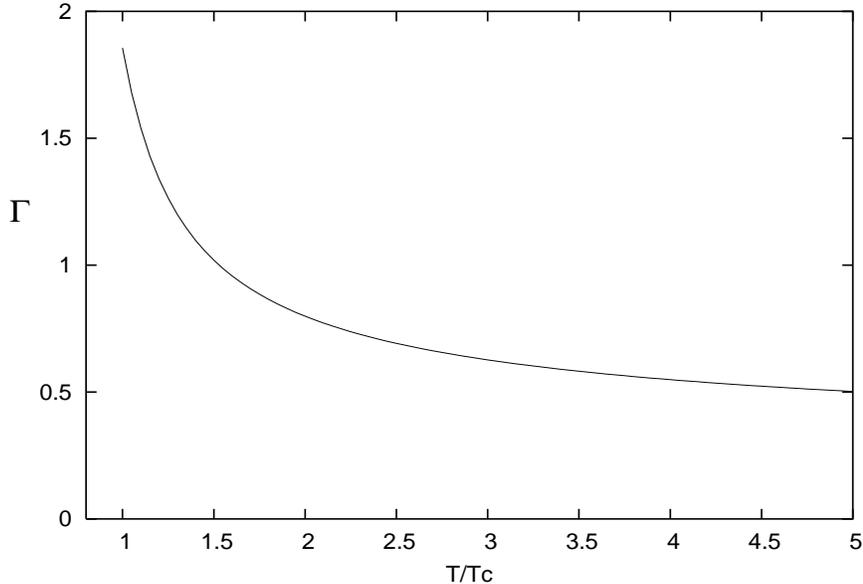}
\caption{ Plot of $\Gamma $ as a function of $T/T_c$ for gluon plasma.} 
\end{figure} 
    
\section{qQGP:}
 
As we discussed earlier, there are varieties of qQGP, starting from the work of Goloviznin and Satz \cite{sa.1}, Peshier {\it et. al.} \cite{pe.1} and many other authors \cite{qqgp.1,nqgp.1,nqgp.2}. Recently, we developed a new qQGP \cite{nqgp.1,nqgp.2} which differs from earlier qQGP \cite{pe.1,qqgp.1} in the derivation of thermodynamic (TD) quantities from statistical mechanics (SM). We start from energy density, a well defined in SM, and derive all other TD quantities using TD relations. Whereas in other models, one starts from pressure \cite{pe.1} and hence TD inconsistency and the need of reformulation of SM \cite{qqgp.1}. In our qQGP model also there are different version \cite{nqgp.1,nqgp.2} because of the way we choose the quasi-parton masses. There are varieties of qQGP models as well, depending on different forms of thermal masses and running coupling constants \cite{qqgp.1}. All models, by adjusting the free parameters, explain very well the LGT results. All these qQGP may be called classical. There is another class of qQGP, called dynamical qQGP, which follows from field theoretical approach. Again, one starts from pressure-partition function relation, but thermodynamically consistent because of the self-consistent $\Phi$ derivable approximation \cite{br.1,br.2}. It reproduces leading order perturbative QCD results at weak coupling limit for gluon plasma and  is successful in fitting the LGT results for $T > 2.5 T_c$ \cite{br.1,br.2}, but one needs to use phenomenological model with 2 or 3 parameters to fit the LGT results \cite{rh.1} for $T < 2.5 T_c$.   

Let us consider our simplest qQGP model where we take the thermal mass of gluons is equal to the plasma frequency and it fits the LGT results very well \cite{nqgp.1,nqgp.2}. We start with the energy density \cite{nqgp.1,nqgp.2} for gluon plasma, 
\begin{equation}
\varepsilon =  \frac{1}{V} \sum_k \frac{\epsilon_k e^{- \beta \epsilon_k }}{1 - e^{- \beta \epsilon_k} } 
\,\,, \label{eq:u} 
\end{equation}
where $\epsilon_k \equiv \sqrt{k^2 + m_g^2}$ and $\beta \equiv \frac{1}{T}$. $m_g$ is the thermal mass which is taken to be equal to the plasma frequency $\sqrt{\frac{4 \pi \alpha_s(T) T^2}{3}}$ with the same $\alpha_s(T)$ as given by Eq. (\ref{eq:as}). We assumed here that the whole thermal energy is used to excite quasi-particles and hence the vacuum energy is taken to be zero. As pointed out in Ref. \cite{nqgp.2}, there is no TD inconsistency in our model with this assumption. From above equation, Eq. (\ref{eq:u}), we get, 
\begin{equation}
e(T) = \frac{15}{\pi ^4} 
\sum_{l=1}^\infty  \frac{1}{l^4} 
\left[ (\frac{m_g\,l}{T})^3 K_1 (\frac{m_g\,l}{T}) 
+  3\, (\frac{m_g\,l}{T})^2 K_2 (\frac{m_g\,l}{T}) \right] \,\,.\label{eq:qqgp}  
\end{equation}
Note that the pressure, in both the models, is obtained from the TD relation,
\begin{equation}
\varepsilon  =  T \frac{\partial P}{\partial T} 
- P \,\, ,  \label{eq:td} \end{equation}
on integration, with one integration constant which we fix to the LGT data at $T = T_c$. The adjustable parameter is only $\Lambda_T$ which we adjust such that we get the best fit to LGT results and $\Lambda_T /T_c = .65$. As an example we plotted $\Delta_q$ in Fig. 1. This model is very successful to explain LGT results on all systems \cite{nqgp.1,nqgp.2}. Thus we see that this simple, single parameter qQGP model explains the LGT results as good as SCQGP in the relevent range of temperature $T_c$ to $5 T_c$. But it has the limitation that it fails to reproduce exact leading order perturbative QCD results at weak coupling limit, eventhough it leads to a term of the order of $g^2$.   
\begin{figure}[h]
\centering
\includegraphics[height=8cm,width=12cm]{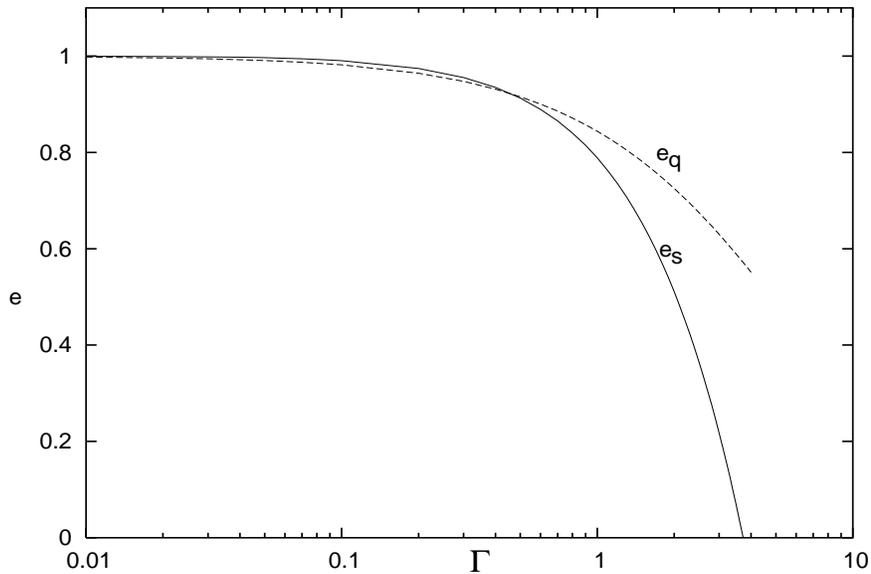}
\caption{ Plot of $e \equiv \frac{\varepsilon}{\varepsilon_{SB}}$ as a function of $\Gamma$ for gluon plasma.} 
\end{figure}

\section{qQGP and SCQGP- Are they same ?:}

It is surprising why both models fit LGT results very well? Is it always possible to model SCQGP using qQGP and vise versa?. We see from Fig. 1 that they almost lie one over other. To address this question we plot $e_s $ and $e_q$ as a function of $\Gamma$ in Fig. 3. This ratio, $e_s(\Gamma)$, for SCQGP is obtained from the Eq. (\ref{eq:scqgp}). The same quantity for qQGP, $e_q(\Gamma)$, may be obtained from Eq. (\ref{eq:qqgp}), by expressing $\alpha_s$ in plasma frequency in terms of $\Gamma$, using Eq. (\ref{eq:ga}). Note that in quasiparticle models the effects of interactions between real partons is replaced by the thermal mass of quasipartons. Therefore, the plasma frequency in our model includes the effects of interactions. Whereas in SCQGP the effect of interactions enter through plasma parameter $\Gamma$. Both plasma frequency and plasma parameter depends on $\alpha_s$, density and/or temperature and hence can be related. Both the models approximately match for $\Gamma < 1$, deviate rapidly for larger $\Gamma$. It is interesting to note, from Fig. 2, that only very close to $T = T_c$, $\Gamma$ is greater than $1$ and hence, except very near to $T = T_c$, both models approximately match. Also note that the EoS of SCQGP, Fig. 3, shows a sign of phase transition, $e_s \rightarrow 0$ for some $\Gamma$, but qQGP does not show such features for any $\Gamma$ and hence qQGP model may not model QGP liquid as claimed by \cite{pe.2}.       
\section{Results and Conclusions:}

We analyzed and compared two different phenomenological models of QGP, quasi-particle model and strongly coupled plasma model. The results are presented in Fig. 1, where, as an example, $\Delta \equiv (\varepsilon - 3 P)/T^4$ of the models were compared with LGT results \cite{bo.1} for gluon plasma and a remarkable good fit was obtained for both the models. Further, plots of the model almost lie one over other. In Fig. 3, the ratio $e(\Gamma) \equiv \frac{\varepsilon}{\varepsilon_{SB}}$ is plotted for both the model as a function of plasma parameter $\Gamma$. We found that SCQGP may be modelled by qQGP and vise versa for $\Gamma < 1$, but models deviate rapidly as a function of $\Gamma$ for larger $\Gamma$. By fitting the LGT results of gluon plasma, both the models predict that $\Gamma$ is always less than $1$, except for $T_c < T < 1.5 T_c$, and hence QGP may be strongly coupled plasma state, rather than liquid or crystals state. For, $T_c < T < 1.5 T_c$, QGP may be in gas-liquid phase transition region which overlaps with deconfinement-confinement phase transition or cross over region which may contain many non-perturbative objects like colored and colorless hadrons \cite{sh.1}, monopoles \cite{ch.1,sh.4} etc. which anyway make the region very blurred. Eventhough QGP is SCQGP one may use fluid theory to describe it's evolution as done in electrodynamic plasma where it is found that about 80 \% properties may be explained. In this sense we can treat QGP as a liquid, as pointed out by Nagle \cite{na.1}. 
Similarly, the use of term sQGP may be a general one, including all models of strong interactions, QCD, but it may be useful to classify them with specific names like SCQGP \cite{ba.1,ba.2} or cQGP \cite{sh.2}, qQGP \cite{sa.1,pe.1,qqgp.1,nqgp.1,nqgp.2}, colored and colorless hadron gas model \cite{sh.1}, supersymmetric model \cite{sh.3}, monopole model \cite{ch.1,sh.4}, so on and compare them.    
\noindent 

{\bf Acknowledgement:} 

I thank the organizer of the "Quark Matter 2008", 20$^{th}$ International Conference on Ultra-Relativistic Nucleus Nucleus Collisions, Jaipur, India (2008), for the financial help to attend the conference, which inspired me to present this result.

\end{document}